\pdfoutput=1

\documentclass[12pt,a4paper]{article} 
\usepackage[top=4cm,bottom=2cm,left=2cm,right=2cm]{geometry}

\usepackage[dvips]{graphicx}
\usepackage{hyperref}
\usepackage{amsmath}
\usepackage{subfigure}

\usepackage{color}
\definecolor{greencolor}{rgb}{0,0.5,0.2}
\definecolor{redcolor}{rgb}{.7,0.,0.}
\definecolor{bluecolor}{rgb}{0,0.,1.}
\definecolor{greycolor}{rgb}{.5,.5,.5}

\begin{document}

\title{{Using Complex Networks to Quantify Consistency in the Use of Words}}

\author{D. R. AMANCIO$^1$, O. N. OLIVEIRA JR.$^1$ and L. da F. COSTA$^1$ }
\date{$^1$  Institute of Physics of S\~ao Carlos \\
	University of S\~ao Paulo, P. O. Box 369, Postal Code 13560-970 \\
	S\~ao Carlos, S\~ao Paulo, Brazil \\}

\maketitle

\begin{abstract}
 {In this paper we quantify the consistency of word usage in written texts represented by complex networks, where words were taken as nodes, by measuring the degree of preservation of the node neighborhood.} Words were considered highly consistent if the authors used them with the same neighborhood. When ranked according to the consistency of use, the words obeyed a log-normal distribution, in contrast to the Zipfs law that applies to the frequency of use. Consistency correlated positively with the familiarity and frequency of use, and negatively with ambiguity and age of acquisition. An inspection of some highly consistent words confirmed that they are used in very limited semantic contexts. A comparison of consistency indices for $8$ authors indicated that these indices may be employed for author recognition. Indeed, as expected authors of novels could be distinguished from those who wrote scientific texts. Our analysis demonstrated the suitability of the consistency indices, which can now be applied in other tasks, such as emotion recognition.
\end{abstract}

\section{Introduction}

Since the dawn of humanity, the ability of communication has been proven to be an essential factor for preservation of life and for the maintenance of social relations. Writing, a major manifestation of communication, whose invention dates back to 3200 B.C., also established itself as one of the key skills developed by mankind. Among its main advantages over the spoken language are the joint capacity of portability and permanence, which ensure that thoughts, ideas, facts and stories are preserved. Despite this ubiquity, the writing skill cannot be considered a trivial or ordinary task. Even after language acquisition, the construction of a concise, precise and well concatenated text requires organized thought, the ability to use expressive linguistic resources and the analytical interpretation of reality.

In addition to the difficulties imposed by the grammar mechanisms in written language~\cite{complex1}, there is a factor related to the semantic level of detail to recreate and interpret the author's original idea. Since it is not possible to specify all the details in a finite piece of text, the author must always focus on the desired level of generality. Thus, if little detail is provided the reader must fill in the blanks using his/her own semantic knowledge and experience about the world. On the other hand, for an excessively detailed text there is no room for inferences, which makes reading more objective. This dichotomy between objectivity and subjectivity has been explored in various ways in different genres of writing. While scientific texts, newspapers, magazines and reports tend to use a more objective approach, literary and artistic texts tend to present themselves subjectively. In both cases, the degree of objectivity (or subjectivity) varies depending on both the text size (long texts tend to be more detailed) and number of descriptive words (a text with many adjectives for instance tend to be very detailed). Equally important seems to be context induced by words~\cite{clustering1,clustering2}, since words inducing restricted contexts somehow limit the ability to extrapolate ideas, and this makes the text more objective.

With this correlation between induction and objectivity as motivation~\cite{motiva}, in this paper we address the problem of quantifying the level of {consistency inherent in words, i.e. the degree of preservation of their neighbors}, to understand the reasons why a word is used in a more or less consistent way. We use the term consistency because words {whose neighbors are preserved} will tend to be used in the same, consistent way by different authors in distinct types of text. Using the concepts and methodologies from complex networks~\cite{costa1,costa2,structure} to analyze the relationship between concepts, we developed a series of indices to measure consistency. These indices are based on the idea that if a word induces a limited set of contexts, then the neighborhood of that word tends to be maintained even in texts written by different writers. In fact, this seems a reasonable assumption, since it is known that syntactically related words also tend to be semantically related. With the indices created using this methodology, we shall show that the distribution of consistency does not follow a power law~\cite{power1,power2}, unlike the case of the frequency of words (Zipf Law)~\cite{zipflaw}. Instead, the distribution seems to follow a log-normal distribution. Furthermore, the greater the familiarity, the number of distinct neighbors and frequency in the language, the more consistent is the word. As for the semantic factors, we showed that consistent words tend to preserve not only the lexical neighborhood, but also the semantic context. Finally, we show how the quantification of consistency can be useful in tasks such as those related to authorship recognition.

\section{Methodology}

\subsection{Dataset}

The distinct contexts in which words are used were investigated with a database comprising several books from the Gutenberg project repository \footnote{http://www.gutenberg.org/}, whose list appears in Table \ref{tab.1}.
Although the number of books differs among authors, the size of the corpus for each author has a fixed size (180,500 tokens). Thus, the difference of the corpora size has little interference in the analysis of consistency of words.

\begin{table}
\centering
\caption{\label{tab.1}Database employed in the experiments.}
\begin{tabular}{|c|c|}
\hline
\textbf{Author}&\textbf{Book}\\
\hline
Arthur Conan Doyle				&	Uncle Bernac - A Memory of the Empire	\\
										&	The Tragedy of the Korosko \\
										&	The Valley of Fear	\\
										&	The War in South Africa \\
										&  The White Company	\\
										&  Through the Magic Door	\\
										&  The Adventures of Sherlock Holmes \\
\hline
Bram Stoker       				&	Dracula's Guest	\\
										&	The Jewel Of Seven Stars \\
										&	The Lady of the Shroud	\\
										&	Lair of the White Worm \\
										&  The Man	\\
\hline
Charles Darwin						&  Coral Reefs \\
										& 	On the Origin of Species by Means of Natural Selection \\
										& 	The Voyage of the Beagle \\
										&  The Different Forms of Flowers on Plants of the Same Species \\
\hline
Charles Dickens					        &  American Notes \\
										&  A Tale of Two Cities \\
										&  Hard Times \\
										&  The Old Curiosity Shop \\
\hline
Thomas Hardy 						&  A Changed Man; and Other Tales \\
										&  Desperate Remedies	\\
										&  Far from the Madding Crowd	\\
										&  The Hand of Ethelberta \\
\hline
Pelham Grenville Wodehouse		&  My Man Jeeves \\
										&  Tales of St. Austin's \\
										&  The Adventures of Sally \\
										&  The Clicking of Cuthbert \\
										&  The Gem Collector \\
										&  The Man with Two Left Feet \\
										&  The Pothunters \\
										& 	The Swoop! \\
										&  The White Feather \\
\hline
Virginia Woolf						&  Jacob's Room \\
										&  Monday or Tuesday \\
										& 	Night and Day \\
										&	The Voyage Out \\
\hline
William Wordsworth				&  Lyrical Ballads, with Other Poems - Volume 1 \\
										&  The Poetical Works of William Wordsworth - Volume 1-3  \\
\hline
\end{tabular}
\end{table}

\subsection{Modeling Texts as Complex Networks}

{Language issues have attracted the interest of many researchers in recent years~\cite{madaga,js1,monte1,js2,njp1,js3,physicaA,js4}. For instance, physicists have used dynamical systems~\cite{barabasi,barryman} and complex networks~\cite{amancio1,sumario,sole1,sole2,macc,review,mendes,co-occur} to study various aspects of language}. Many of these studies have used the co-occurrence model~\cite{mendes,co-occur,amancio2} to link adjacent words in the text. The idea behind this modeling comes from the use of co-occurrence statistics at various scales, from bigram statistics to discourse scale windows, which has been widespread in document analysis and retrieval for at least two decades~\cite{japan,landauer,tam,bigrams}. Because we are interested in the neighborhood properties of words to examine the preservation of induced contexts, we chose to use this model, which is described as follows.

\begin{figure}
\begin{center}
	\includegraphics[width=0.7\textwidth]{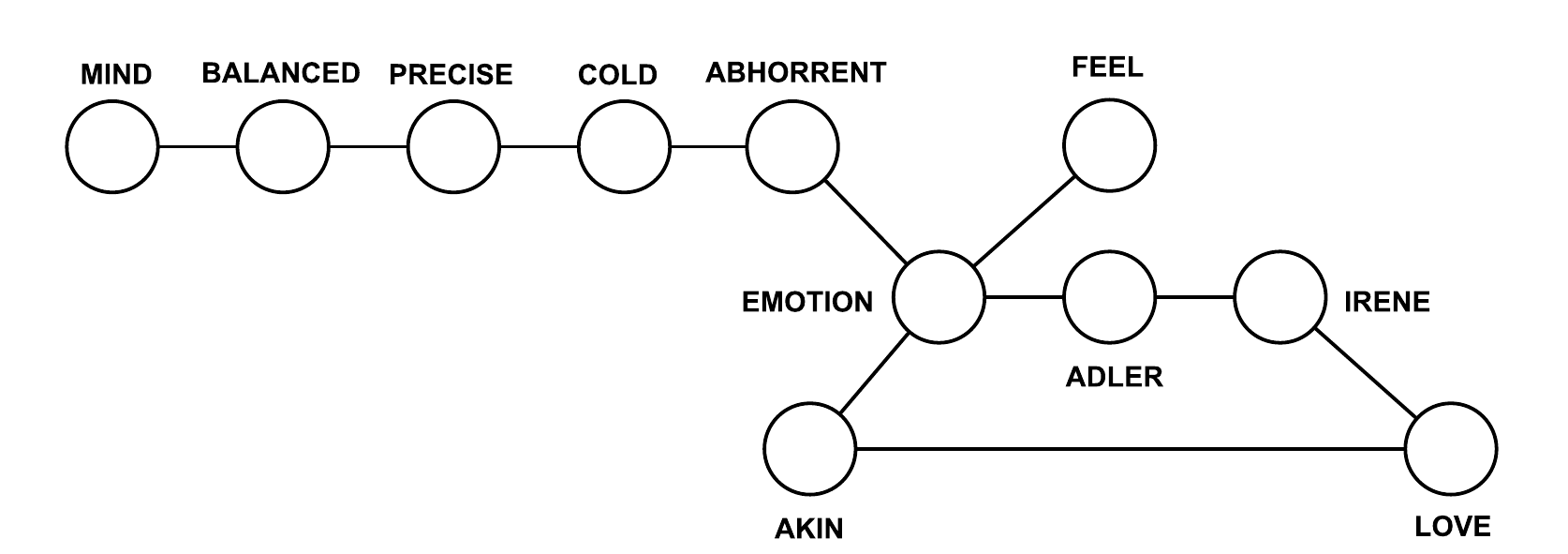}
\end{center}
\caption{Example of network obtained for the extract: {\it ``It was not that he felt any emotion akin to love for Irene Adler. All emotions, and that
one particularly, were abhorrent to his cold, precise but admirably balanced mind''} of the book {\it The Adventures of Sherlock Holmes}, by Arthur Conan Doyle.}
\label{fig.1}
\end{figure}

The modeling procedure started with a pre-processing step, where stopwords (i.e., words with little semantic meaning), such as articles and prepositions, were removed from the text. Although the frequency of such words may be useful in distinguishing writers' personal characteristics~\cite{rec1} we decided to ignore them because we are only interested in the contextual semantic preservation. The remaining words were then lemmatized so that conjugated verbs and nouns in the plural form were converted respectively to their infinitive and singular forms. Thus words related to the same concept but with distinct inflections were taken as a single node in the network. This lemmatization was performed with the MXPOST part-of-speech tagger~\cite{ratim}, based on a maximum entropy model~\cite{maxentr}. This was done to resolve ambiguities during the conversion to the canonical form (infinitive and plural). After this preprocessing, each distinct word became a node and the neighborhood relationship between words defined the set of edges. To illustrate the procedure, we created a small network for the following text extract: {\it ``It was not that he felt any emotion akin to love for Irene Adler. All emotions, and that one particularly, were abhorrent to his cold, precise but admirably balanced mind.''} obtained from the book {\it The Adventures of Sherlock Holmes}, by Arthur Conan Doyle. Table \ref{tab.2} summarizes the pre-processing steps and Figure \ref{fig.1} displays the resulting network.

\begin{table}
\centering
\caption{\label{tab.2} The pre-processing step involves two procedures: (i) Removal of stopwords and (ii) Lemmatization of the remaining words. }
\begin{tabular}{|c|c|c|}
\hline
\textbf{Original} & \textbf{Without stopwords}  & \textbf{Lemmatized}  \\
\hline
It was not that he felt any emotion & felt emotion          &	feel emotion \\
akin to love for Irene Adler.			& akin love Irene Adler &	akin love Irene Adler \\
All emotions, and that one 			& emotions					& 	emotion	\\
particularly, were abhorrent to his & abhorrent					&	abhorrent \\
cold ,precise but admirably 			& cold precise 			&  cold precise \\
balanced mind.								& balanced mind			&  balanced mind \\
\hline
\end{tabular}
\end{table}

\begin{figure}
\begin{center}
	\includegraphics[width=0.5\textwidth]{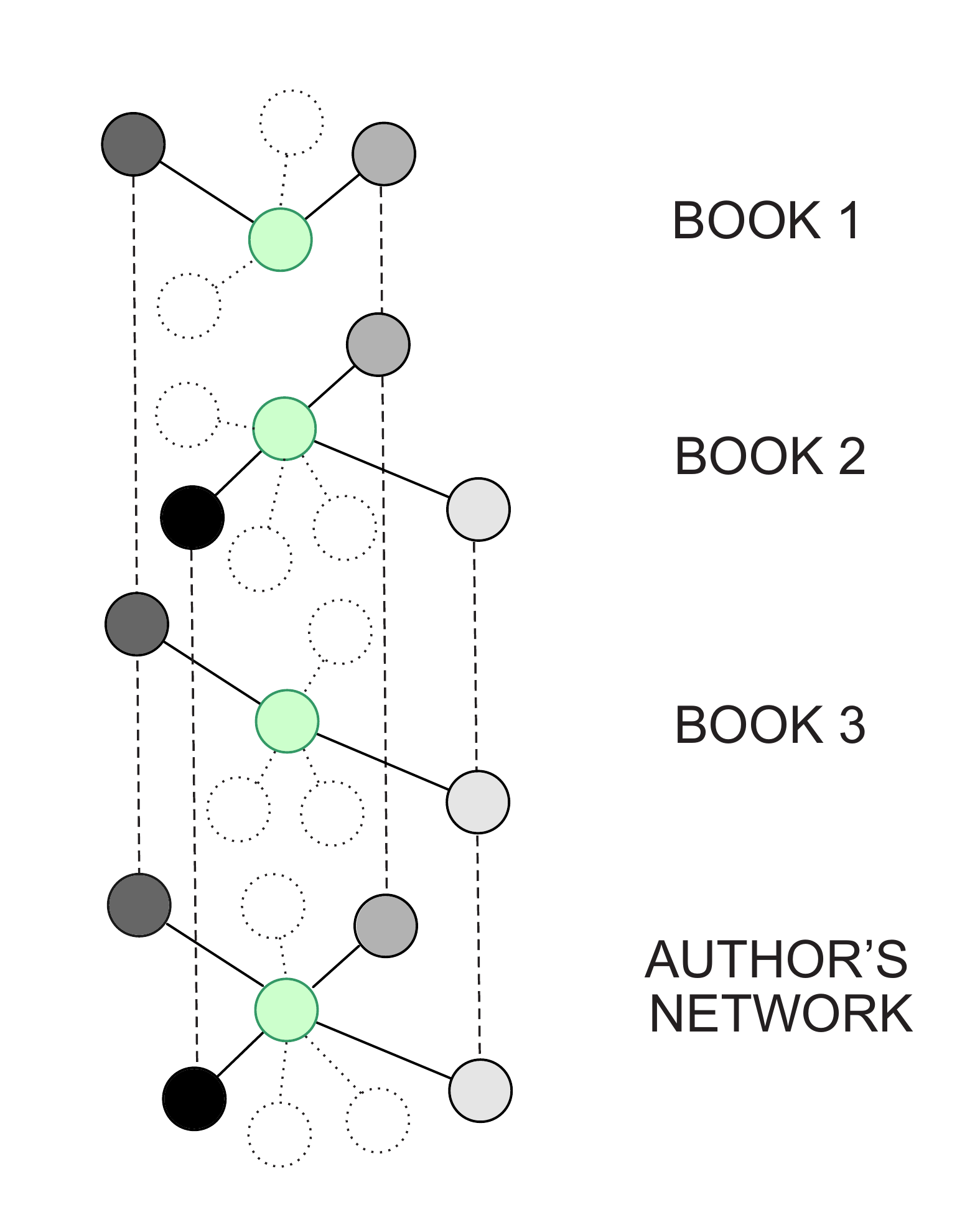}
\end{center}
\caption{Example of derivation of a specific network for an author from the network of books. Note that if a neighbor appears in at least one of the networks of books, it will also appear on the author's network.}
\label{fig.2}
\end{figure}

The networks for each of the books of an author were then joined to obtain the so-called author's network, reflecting the association of words generated by that author. That is to say, if a given node appears in one of the network of books, then it will also appear in the author's network. Similarly, two vertices were connected in the author's network if both appeared connected at least in one network of the books. The derivation process is illustrated in Figure \ref{fig.2}.

\subsection{Consistency Indices} \label{cis}

In this section we describe the indices proposed to measure the consistency $\mathcal{C}$ with which words are used. Because obtaining $\mathcal{C}$ demands that the neighborhood of each word for each author is known, it was calculated only for the $2,880$ words that appeared in the networks for all authors.

\subsubsection{$\mathcal{C}$ Based on the Histogram of Co-occurrence of Neighbors}

To derive this index for a given word $w$, consider the vector $\overrightarrow{v^w}$ storing the frequency of each of the $k(w)$ distinct neighbors of $w$ over all authors' networks. Hence, $v_i^w$ stores the number of authors for which a given neighbor $i$ appeared connected to the word $w$. The element $v_i^w$ of $\overrightarrow{v^w}$ ranges from 1 to 8, since $n$ = 8, the number of  authors' studied. The consistency based on $\overrightarrow{v^w}$ increases with the sum of the vector components, since high $v_i^w$ means that the corresponding association of words is repeated by several authors. The consistency index is calculated as:

\begin{equation} \label{cons1}
	\mathcal{C}_{hist}(w) = \frac{ s(w) - k(w) }{(n-1)k(w)} = \frac{ s(w) - k(w) }{7k(w)},
\end{equation}
where $s$ is given by:
\begin{equation}
	s(w) = \sum_{i=1}^{k(w)} v_i^w
\end{equation}

	Note that while the numerator in eq. (\ref{cons1}) represents the sum of $v_i^w$ derived from the minimum possible sum (which occurs when each of the $k$ neighbors occurs in only one of the authors' networks), the denominator is the range of the sum (the largest sum occurs when every neighbor appears in all $n$ networks). Consequently, $\mathcal{C}_{hist}$ ranges between 0 and 1.

\subsubsection{$\mathcal{C}$ Based on the Cosine Similarity}

	Consistency may be calculated for words considering two distinct authors' networks $A^{i}$ and $A^j$, comprising the set of nodes denoted by $V_i$ and $V_j$ respectively. Initially, $V_i$ and $V_j$ were expanded (keeping the original links) so that the new set of vertices $V_{ij}$ = $V_i$ $\cup$ $V_j$. Let $w$ be the word whose consistency is being calculated. The number of neighbors of $w$ appearing in both $A^i$ and $A^j$ is:

\begin{equation} \label{comum}
	\mathcal{N}_{ij}(w) = \sum_{k} A^i_{wk} A^j_{wk}
\end{equation}

In spite of capturing the number of neighbors in common, it is difficult to interpret if the value of $\mathcal{N}_{ij}$ is high or low, since it is not normalized. We have therefore normalized equation \ref{comum} dividing it by the geometric mean of the degrees of the word $w$ in both networks. Thus, if $k^i_w$ and $k^j_w$, given by
\begin{equation} \label{grau}
	k_w = \sum_{x} A_{wx}
\end{equation}
represent the degrees of the word $w$ respectively in $A^{i}$ and $A^j$, then the normalized number of shared neighbors is:
\begin{equation} \label{cosine}
	\mathcal{N}_{ij}'(w) = \frac{\sum_{k}  A^i_{wk} A^j_{wk}}{\sqrt{k^i_w k^j_w}}
\end{equation}

	In order to take into account all pairs of authors in computing the consistency, we defined consistency as the average of $\mathcal{N}_{ij}'$ over the $n$($n$-1)$/$2 pairs of distinct networks:

\begin{equation}
	\mathcal{C}_{cos}(w) = \frac{2}{n(n-1)} \sum_{i=1}^{n} \sum_{j=i+1}^{n} \mathcal{N}_{ij}'(w)
\end{equation}

	It is also worth mentioning that $\mathcal{C}_{cos}$ ranges between 0 and 1, analogously to $\mathcal{N}_{ij}'$. Actually, $\mathcal{C}_{cos}$ can be interpreted as the cosine of the angle between the vectors whose elements indicate the presence or absence of a particular word as neighbor. Thus, if these vectors are similar, the angle between them is small and the cosine is high, indicating high consistency.

\subsubsection{$\mathcal{C}$ Based on the Pearson Correlation Coefficient}

One can estimate whether $\mathcal{N}_{ij}$ in eq. (\ref{comum}) is high by comparing it with the expected number of neighbors assuming a random choice of neighborhood. The expected number of common neighbors for word $w$ in the networks $A^{i}$ and $A^j$ is given by $k^i_w$ and $k^j_w$, as defined in equation \ref{grau}. If $w$ in the network $A^{i}$ randomly chooses one of its  $k^i_w$ neighbors, then the probability of choosing a node which is also a neighbor of $w$ in $A^j$ would be $k^j_w/N$. Applying this reasoning for the $k^i_w$ neighbors of $w$ in $A^i$, the expected number of common neighbors is $k^i_w k^j_w / N$, where $N$ represents the number of nodes in the network. Therefore, if $\mathcal{N}_{ij} > k^i_w k^j_w / N$, consistency is higher than expected. Thus, defining consistency as the difference between the actual and expected numbers of neighbors in common, we obtain $\mathcal{C}_{r}$:

\begin{eqnarray} \label{pearson}
	\mathcal{C}_{r}(w) &=& \mathcal{N}_{ij} - \frac{k^i_w k^j_w}{N} = \sum_{k}  A^i_{wk} A^j_{wk} - \frac{k^i_w k^j_w}{N} \nonumber \\
			&=& \sum_{k} A^i_{wk} A^j_{wk} - N  \overline{k}^i_w  \overline{k}^j_w \nonumber \\
			&=& \sum_{k} \Big{(}   A^i_{wk} A^j_{wk} - \overline{k}^i_w  \overline{k}^j_w \Big{)} \nonumber \\
			&=& \sum_{k} \Big{(} A^i_{wk} -  \overline{k}^i_w \Big{)} \Big{(}  A^j_{wk} -  \overline{k}^j_w \Big{)},
\end{eqnarray}
where the notation $\overline{k}$ represents the degree normalized by the number of nodes in the network:
\begin{equation}
	\overline{k}_w = \frac{1}{N} \sum_{x} A_{wx}.
\end{equation}
Eq. (\ref{pearson}) can be interpreted as a covariance, i.e., a non-normalized correlation. To transform $\mathcal{C}_{r}$ into a normalized measurement, $\mathcal{C}_{r}$ is divided by the corresponding standard deviations of the vectors $A^i_{wk}$ and $A^j_{wk}$, $k$ = 1 .. $n$. The covariance then becomes the Pearson correlation coefficient $r_{ij}$:

\begin{equation} \label{finalpearson}
	\mathcal{C}_{r}'(w) = r_{ij} = \frac{\sum_{k} \Big{(} A^i_{wk} -  \overline{k}^i_w \Big{)} \Big{(}  A^j_{wk} - \overline{k}^j_w \Big{)}}{ \sqrt{\sum_{k} \Big{(}A^i_{wk} -  \overline{k}^i_w \Big{)}^2} \sqrt{\sum_{k} \Big{(} A^j_{wk} -  \overline{k}^j_i \Big{)}^2} }
\end{equation}

	With consistency defined as in eq. (\ref{finalpearson}), $\mathcal{C}_{r}'$ ranges between -1 and 1. In order to restrict the range between zero and 1, the following linear transformation was performed in $\mathcal{C}_{r}'$, deriving $\mathcal{C}_{r}''$:

\begin{equation}
	\mathcal{C}_{r}'' = \frac{\mathcal{C}_{r}' + 1}{2}
\end{equation}

Now, if $\mathcal{C}_{r}'' > 0.5$, then the quantity of shared neighbors is greater than the expected by chance.

\subsubsection{$\mathcal{C}$ based on the Leicht-Holme-Newman Index~\cite{lht}}

In the consistency index based on the Pearson correlation coefficient, $\mathcal{C}_{r}$ derived from the difference between the number of common neighbors found and the number of common neighbors expected by chance. For the Leicht-Holme-Newman Index~\cite{lht}, the new measure of consistency is obtained from the ratio between these numbers, as follows:

\begin{equation}
	\mathcal{C}_{LHN} = \frac{\mathcal{N}_{ij}(w)}{k^i_w k^j_w} = N \frac{\sum_{k} A^i_{wk} A^j_{wk}}{\sum_{k} A^i_{wk} \sum_{k} A^j_{wk}}
\end{equation}

The important threshold now is the number 1, since for $\mathcal{C}_{LHN}$ above 1, the consistency is higher than expected.

\subsubsection{$\mathcal{C}$ Based on the Sorensen Index~\cite{sorensen}}

Unlike the metrics described above, this measure is based on the dissimilarity of the set of neighbors of a given word in two networks. The vectors $A^{i}_w$ and $A^{j}_w$, which store the neighborhood of the word $w$, are used again, with the dissimilarity of neighbors being computed as the corresponding squared normalized Euclidean distance:

\begin{eqnarray}
	\mathcal{C}_{euc} &=& \frac{\sum_{k} \Big{(} A^i_{wk} - A^j_{wk} \Big{)} ^ 2}{ k^i_w  +  k^j_w } = \frac{\sum_{k} \Big{(} A^i_{wk} + A^j_{wk} - 2 A^i_{wk}  A^j_{wk} \Big{)} }{k^i_w  +  k^j_w} \nonumber \\
	&=& 1 - 2 \frac{\mathcal{N}_{ij}(w)}{k^i_w  +  k^j_w }
\end{eqnarray}

Since the distance between the vectors was divided by the maximum possible distance (given by $k^i_w  +  k^j_w$), $\mathcal{C}_{euc}$ ranges between 0 and 1. To make this measure consistent with other metrics, we converted it from a dissimilarity to a similarity measure with the following transformation:

\begin{eqnarray}
	\mathcal{C}_{s} &=& 1 - \mathcal{C}_{euc} = 2 \frac{\mathcal{N}_{ij}(w)}{k^i_w  +  k^j_w }
\end{eqnarray}

\subsubsection{$\mathcal{C}$ based on the Frequency of the Shared Neighbors}

In this index, we also consider the frequency with which a given neighbor was employed by each author. This new measure is justified because neighbors may appear with quite different frequencies. For example, if a given association of words occurs 100 times for an author and only once for another author (considering texts of the same size), the consistency index would still be maximum according to the indices described so far. However, this combination of words is clearly not consistent, since the frequencies are quite different.

The disparity in frequency is considered as follows. Suppose that word $w$ has $N$ distinct neighbors in the corpus of $n$ authors. Let $\upsilon_k$ be one of the neighbors of $w$. If $\upsilon_k$ appears associated to $w$ $f_k^i$ times in $A^i$ and $f_k^j$ times in $A^j$, then the consistency of $w$ regarding the association $w \leftrightarrow \upsilon_k$ is:

\begin{equation}
	\mathcal{C}_{f}^{ij}(w \leftrightarrow \upsilon_k) = 1 - \frac{|f_k^i - f_k^j|}{f_k^i + f_k^j}
\end{equation}

	To consider all the neighbors, $\mathcal{C}_{f}^{ij}(w)$ is computed as an average over $\mathcal{C}_{f}^{ij}(w \leftrightarrow \upsilon_k)$, provided that $f_k^i + f_k^j > 0$. Assuming that this condition occurs $t$ times, $\mathcal{C}_{f}^{ij}(w)$ is:

\begin{equation}
	\mathcal{C}_{f}^{ij}(w) = \frac{1}{t} \sum_{\substack{k \\ f_k^i + f_k^j > 0}} \mathcal{C}_{f}^{ij}(w \leftrightarrow \upsilon_k)
\end{equation}

\section{Results and Discussion}

\subsection{Analysis of distribution and inter correlation of consistency indices}

\begin{figure}
\begin{center}
	\includegraphics[width=0.75 \textwidth]{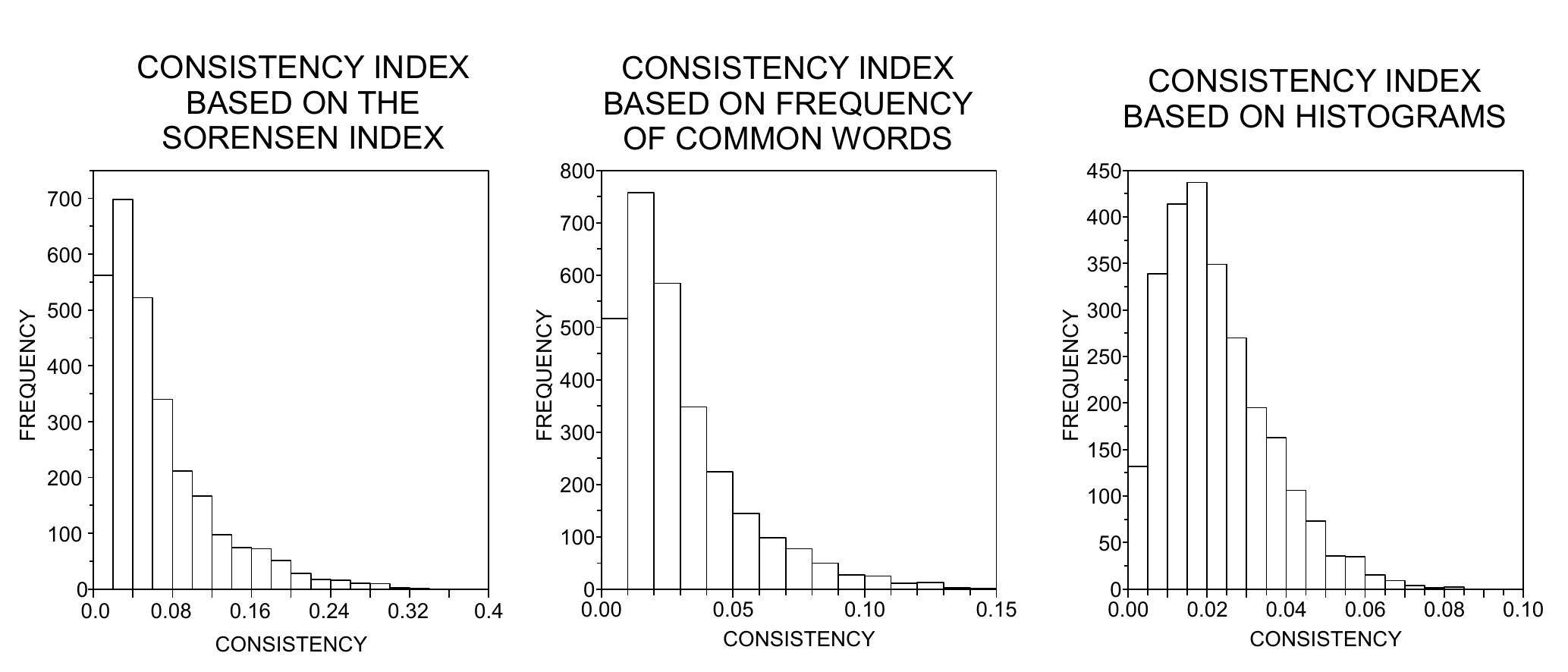}
\end{center}
\caption{Distribution of consistency for 2,880 words. Regardless of the definition employed to quantify consistency, a similar distribution was obtained.}
\label{fig.3}
\end{figure}

The consistency indices defined in Section \ref{cis} were first used to examine the distribution of consistency for the 2,880 words under analysis. The distributions display a peak and two asymmetric tails for all the indices used, as illustrated in Figure \ref{fig.3} for the Sorensen index and the indices based on the frequency of words in common and on histograms. One infers that it is rare for a word to take extreme consistency values (low or high consistency), but it is far more rare for a word to take very high consistency values. Formally, this finding is confirmed by the log-normal distribution that is obeyed by the indices. In all cases, the data could be explained by a log-normal function

\begin{equation}
	f(x;x_c,\sigma,\mathcal{A}) = \frac{\mathcal{A}}{\sqrt{2\pi} \sigma x} e^{-\frac{(\ln x/x_c)^2}{2 \sigma ^ 2}},
\end{equation}
where $x_c$, $\sigma$ and $\mathcal{A}$ correspond to the free parameters of the distribution. Table \ref{tab.25} summarizes the parameters for each case, and confirms the suitability of the fitting with Pearson-squared R$^2$ $\sim$ 1 and chi-square $\chi^2$ $\ll$ 1.

Log-normal distributions are largely found in non-linguistic contexts (see e.g.~\cite{ref0,ref3,ref4,ref5,ref6,ref7}), but it is less common in linguistics, for which statistical distributions such as the Zipf's law~\cite{zipflaw} prevail, as in the case of frequency and ranking of words~\cite{power1}. Nevertheless, log-normal distributions have been reported for random variables in natural language issues. For example, Williams~\cite{williams} showed that the length of sentences seems to follow a log-normal distribution. Similarly, Herdan~\cite{herdan} found that the length of spoken words in phone conversation also follows this distribution. {Log-normal distributions are usually generated by processes following proportionality laws~\cite{Gibrata,Gibratb,orig}. For this reason, the consistency can be thought as a result of a growth process governed by the $\alpha_k$ constant, which also follows a log-normal distribution. To verify why this statement is true, suppose that a given word is initially used by only $2$ authors with consistency $\mathcal{C}_0$. For each new writer who uses this word, the current consistency increases or decreases according to the factor $\alpha_k$:}
\begin{equation}
    \mathcal{C}_k = \alpha_k ~ \mathcal{C}_0
\end{equation}
{After $n+2$ authors use the word, the final consistency $\mathcal{C}_n$ will be given by:}
\begin{equation}
    \mathcal{C}_n = \mathcal{C}_0 \prod_{i=1}^{n} \alpha_i
\end{equation}
{Actually, the consistency of the word is quantified as a percentage of the current consistency with each new use and this percentage is independent of the consistency currently observed. Since we assume that $\alpha_k$ is {log-normally} distributed, then $\mathcal{C}_k$ will also follow a log-normal distribution, because the product of log-normal distributions also generates a log-normal distribution~\cite{orig}. The requirement that $\alpha_k$ follows a log-normal distribution can be disregarded if we consider that many authors use the word. Because $\ln \mathcal{C}_n$ is given by:}
\begin{equation}
    \ln \mathcal{C}_n = \sum_{i=1}^{n} \ln \alpha_i + \ln \mathcal{C}_0
\end{equation}
{and since according to the central limit theorem $\sum_{i=1}^{n} \alpha_i$ follows a normal distribution, it is possible to state that  $\ln \mathcal{C}_n$ is  normally distributed and then $\mathcal{C}_n$ follows a log-normal distribution.}

\begin{table}
\caption{\label{tab.25} Parameters obtained from fitting the consistency probability density functions. All distributions follow a log-normal distribution, since R$^2$ $\sim$ 1 and $\chi^2$ $\ll$ 1.}
\begin{tabular}{|c|c|c|c|c|c|}
\hline
$\mathcal{C}$ & $x_c$  & $\sigma$	&	$\mathcal{A}$ &	R$^2$	& $\chi^2$ \\
\hline
Frequency	&	0.02416 $\pm$ 0.0011 &	0.97641 $\pm$ 0.03091	& 0.00542 $\pm$ 0.00014	& 0.991 & 1.53 $\cdot$ $10^{-5}$ \\
Histogram	&	0.02389 $\pm$ 0.0018	&	0.89291 $\pm$ 0.04733	& 0.00342 $\pm$ 0.00019 & 0.972 & 2.68 $\cdot$ $10^{-5}$ \\
Sorensen		&	0.05076 $\pm$ 0.0020 &	1.04088 $\pm$ 0.02737	& 0.01055 $\pm$ 0.00022 & 0.994 & 8.25 $\cdot$ $10^{-6}$ \\
Pearson 		&	0.04134 $\pm$ 0.0037 &  1.11074 $\pm$ 0.04754	& 0.00683 $\pm$ 0.00036 & 0.986 & 1.49 $\cdot$ $10^{-6}$ \\
Cosine		&	0.06411 $\pm$ 0.0052 &	1.08812  $\pm$ 0.04677	& 0.01232 $\pm$ 0.00058	& 0.989 & 1.52 $\cdot$ $10^{-5}$ \\
LNH			&	3.97112 $\pm$ 0.0536	&	0.65551$ \pm$ 0.01073 & 0.90414 $\pm$	 0.01281	&	0.994	&	1.17 $\cdot$ $10^{-5}$ \\
\hline
\end{tabular}
\end{table}

The similar behavior for the consistency indices in Figure \ref{fig.3} may mean that the indices are correlated. Indeed, the Pearson correlation coefficients ($r$)~\cite{pearson} between pairs of indices were all close to 1 (results not shown), with the exception of the LHN index. The correlation ranged from 0.902 for $\mathcal{C}_r$ and $\mathcal{C}_{hist}$ to 0.996 for  $\mathcal{C}_{cos}$ and $\mathcal{C}_{s}$. Hence, 5 of the indices are equivalent and can be used interchangeably. As for LHN, it is weakly correlated with the other 5 indices, as shown in the first column of Table \ref{tab.26}. Even when the correlation of ranks is used, low values were observed as shown in the second column of Table \ref{tab.26}, which summarizes the values the Spearman's rank correlation coefficient ($\rho$)~\cite{spp}. Therefore, for quantifying consistency in texts in future works, it suffices to consider the measures based on the difference and on the ratio between the number of common neighbors and the expected number of neighbors by chance. For the sake of completeness, however, we chose to analyze all the indices in the next sections.





\begin{table}
\centering
\caption{\label{tab.26} Pearson ($r$) and Spearman ($\rho)$ correlations between LHN and other consistency indices. In both cases, the correlations were low.}
\begin{tabular}{|c|c|c|}
\hline
$\mathcal{C}$ & $r$  & $\rho$ \\
\hline
Histogram	&	0.044	&	0.166	\\
Pearson		&	0.318	&	0.347	\\
Cosine		&	0.301	&	0.315	\\
Sorensen		&	0.293	&	0.295	\\
\hline
\end{tabular}
\end{table}

\subsection{Analysis of Correlation Between Consistency and Linguistic Features}

\begin{figure}
\begin{center}
	\includegraphics[width=0.8 \textwidth]{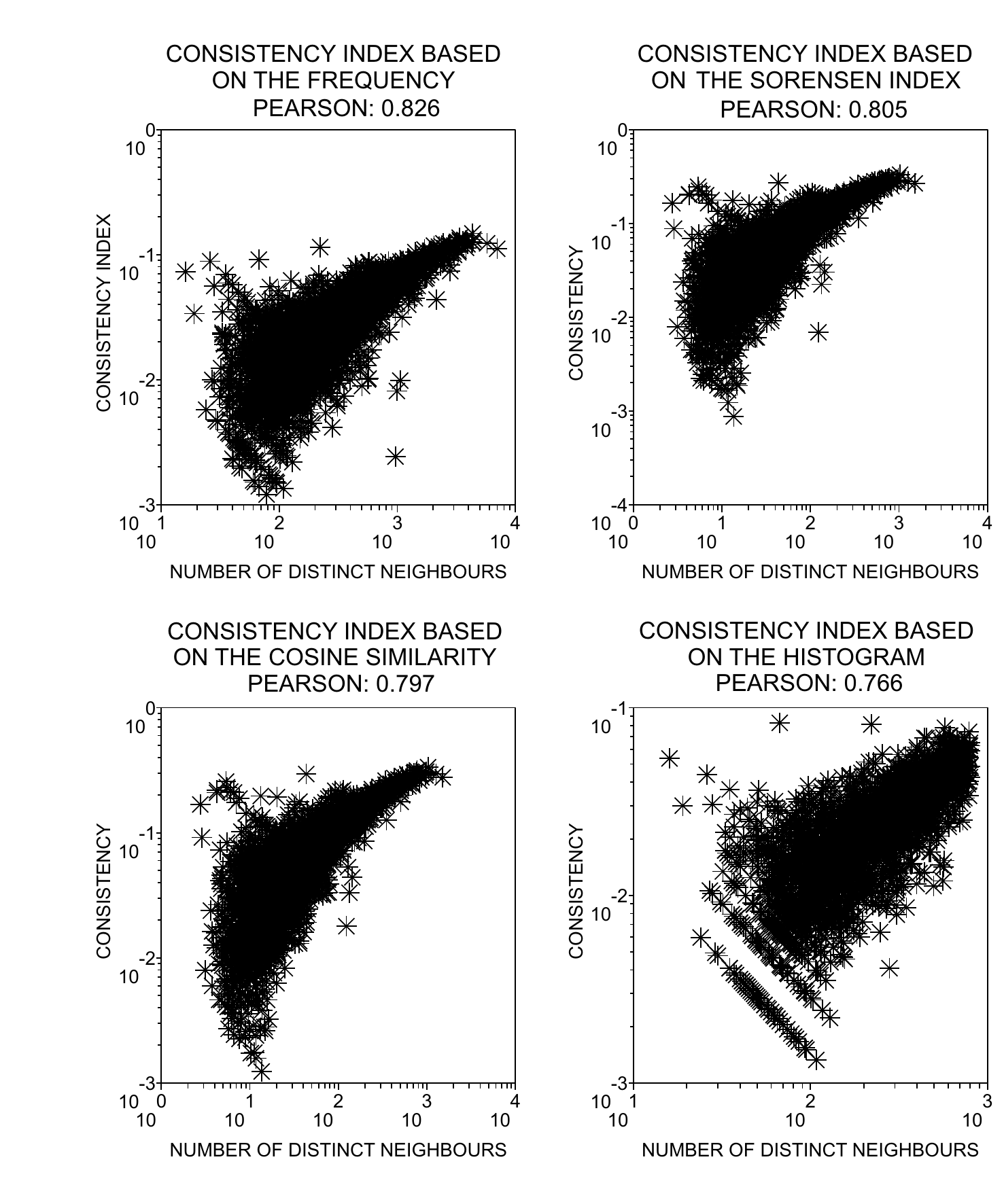}
\end{center}
\caption{Correlation between the number of distinct neighbors and the corresponding consistency. Regardless of the consistency index, there is strong correlation between consistency and the number of neighbors.}
\label{fig.4}
\end{figure}

To understand how consistency measured according to the indices proposed is related to other linguistic factors, we examined the relationship between consistency and the number of distinct neighbors of the word. The scatter plots for 4 consistency indices are given in Figure \ref{fig.4}, indicating a strong positive correlation between the number of distinct neighbors and consistency for all the indices (the indices not shown in Figure \ref{fig.4} exhibit the same behavior). Words with larger number of neighbors in the networks tend to be more consistent. It is possible that words with many neighbors (probably frequently used) are more consistent because they are used in a more uniform fashion as they are widely employed by different writers. Conversely, words with fewer neighbors are probably less frequent words that are more prone to the influence of the writer's personal experiences.

Since the distribution of number of neighbors for the $2,880$ words analyzed is very broad, we divided them into four classes ($C_A$, $C_B$, $C_C$ and $C_D$) according to the number of neighbors as depicted in Table \ref{tab.3}. The correlation between consistency and the following features was calculated \footnote{The quantities (i)-(iv) were obtained from the MRC Psycholinguistic Database~\cite{psycho}.}: (i) age of acquisition (i.e., the age at which a child begins to use the word as part of his/her spoken vocabulary), (ii) familiarity, (iii) imaginability, (iv) frequency of use in the English language; and (v) ambiguity (i.e., the number of distinct meanings extracted from the WordNet~\cite{wordnet}). The correlations for each class and the corresponding consistency that provided the strongest correlations are shown in Table \ref{tab.4}.

\begin{table}
\centering
\caption{\label{tab.3} Number of words $\mathcal{F}$ in each group and range of the number of distinct neighbors $\eta$. While class A comprises words with few distinct neighbors (up to 150), class D comprises words with many distinct neighbors.}
\begin{tabular}{|c|c|c|}
\hline
\textbf{Class} & \textbf{$\eta$}  & \textbf{$\mathcal{F}$}  \\
\hline
Class A	&	0   $\leq \eta \leq$ 150	& 	1,070	\\
Class B	&	151 $\leq \eta \leq$ 300	&	791	\\
Class C 	&	301 $\leq \eta \leq$ 500	&	463	\\
Class D	&	501 $\leq \eta \leq$ 800	& 	256	\\
\hline
\end{tabular}
\end{table}

\begin{table}
\centering
\caption{\label{tab.4} Correlation obtained by comparing linguistic features and consistency of words for classes $C_A$, $C_B$, $C_C$ and $C_D$.}
\begin{tabular}{|c|c|c|}
\hline
\textbf{Linguistic Measure} & \textbf{Correlation for $C_A$} &  \textbf{Consistency Index}  \\
\hline
Age of acquisition	&	-0.09	&	Histogram	 \\
Familiarity				&	+0.18	&	LHN			 \\
Imaginability			&	+0.07	&	Pearson		 \\
Frequency				&	+0.12	&	Sorensen		 \\
Ambiguity				&	-0.13	&	LHN			 \\
\hline
\textbf{Linguistic Measure} & \textbf{Correlation for $C_B$} &  \textbf{Consistency Index}  \\
\hline
Age of acquisition	&	 -0.15 &	Histogram \\
Familiarity				&	 +0.27 &	Histogram \\
Imaginability			&	 +0.13 &	LHN		 \\
Frequency				&	 +0.13 &	Sorensen	 \\
Ambiguity				&	 -0.13 &	LHN		 \\
\hline
\textbf{Linguistic Measure} & \textbf{Correlation for $C_C$} &  \textbf{Consistency Index}  \\
\hline
Age of acquisition	&	-0.24		&	Pearson	 \\
Familiarity				&	+0.37		&	Histogram \\
Imaginability			&	+0.14		&	Sorensen	 \\
Frequency				&	+0.03		&	Cosine	 \\
Ambiguity				&	-0.14		&	LHN		 \\
\hline
\textbf{Linguistic Measure} & \textbf{Correlation for $C_D$} & \textbf{Consistency Index}  \\
\hline
Age of acquisition	&	-0.20	 	&	Sorensen		\\
Familiarity				&	+0.43		&	Sorensen		\\
Imaginability			&	+0.07		&	Histogram	\\
Frequency				&	+0.20		&	Sorensen		\\
Ambiguity				&	-0.20		&	LHN			\\
\hline
\end{tabular}
\end{table}

The only case where the correlation was high occurred for familiarity in classes $C_C$ and $C_D$, with the other correlations being below 0.3. Notwithstanding, interesting trends can be inferred from the sign of the correlations. For example, the negative correlation with the age of acquisition suggests that less consistent words take longer to be learned. This should be expected since a heterogeneous use of weakly consistent words makes them more difficult to be acquired. An analogous reasoning applies to the familiarity and imaginability (ability to visualize a concept), as familiar words tend to induce the same concepts (i.e., well-known words usually bring to mind the same concepts). As for the frequency in the language, the positive correlation indicates that widely used words are more consistent. Indeed, the widespread use of a word probably causes it to be used in a more homogeneous way. Finally, the ambiguity of the word correlated negatively with consistency. This result was also expected, since if a word is ambiguous then it can appear in multiple contexts, with its neighborhood tending to be more heterogeneous.

\begin{figure}
\begin{center}
	\includegraphics[width=0.75\textwidth]{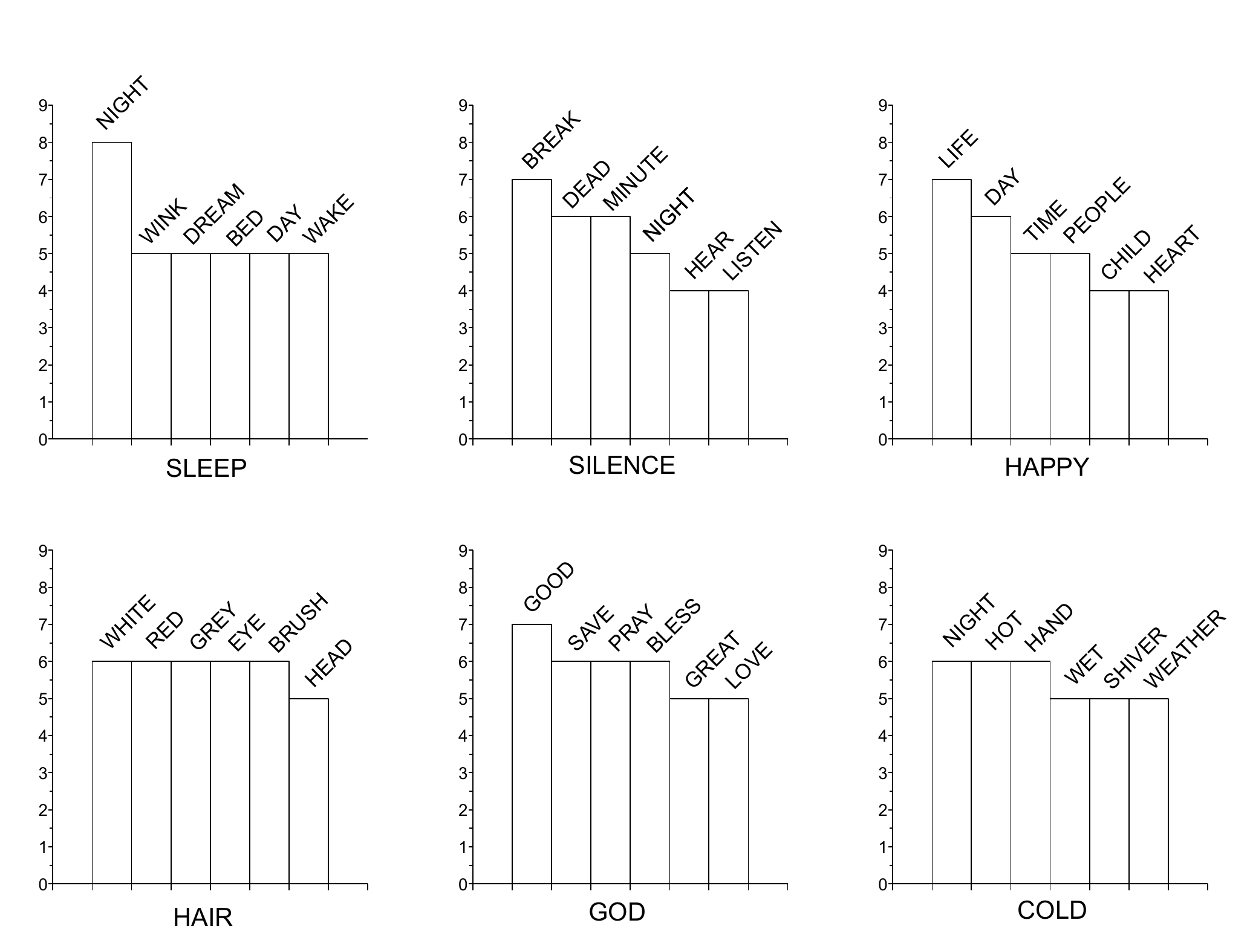}
\end{center}
\caption{Frequency of association of neighbors for the words {\it sleep}, {\it silence}, {\it help}, {\it har}, {\it God} and {\it cold}. Note that for the words shown in this figure (with high values of consistency), their neighborhood seems to be restricted to a single context.}
\label{fig.5}
\end{figure}

In an attempt to understand why some words are more consistent than others, we examined the neighborhood of words possessing the highest and the lowest consistency values. By way of illustration, let us analyze the following strongly consistent words in $C_D$: {\it sleep}, {\it silence}, {\it happy}, {\it hair}, {\it God} and {\it cold}. For each neighbor of each one of these words, we counted the number of authors that associated them. The histograms with the frequencies in Figure \ref{fig.5} indicate that these highly consistent words tend to have semantically related neighbors. For instance, all the 8 authors associated {\it sleep} to {\it night}), {\it wink} and {\it dream}. To confirm this hypothesis of semantic proximity, we compared the neighbors for the texts written by the 8 authors with the neighbors of the semantic network derived from the Edinburgh Associative Thesaurus (EAT)~\cite{eatref}, which connects semantically related concepts. For each of the selected words with high consistency values, we computed its distance to the neighbor of the authors' network in the EAT network. Then, we calculated the ratio between the number of neighbors that are at a specific distance $d$ and the expected number of neighbors that are at the same distance, assuming that the nodes were randomly chosen in the {\it Edinburgh Associative Thesaurus} network. The results in Table \ref{tab.5} show that the number of neighbors immediately related in the EAT network is much higher than expected (which would give a ratio = 1), thus confirming that consistency is related to the limitation of semantic context.



\begin{table}
\centering
\caption{\label{tab.5} Ratio between the number of neighbors of the authors' networks which are at a distance $d$ from the central concept in the EAT network and the expected number of neighbors at a distance $d$ assuming random associations. Interestingly, the immediate neighborhood of the most consistent words is also concentrated in the immediate neighborhood ($d$ = 1) of the EAT networks. It turns out that consistent words induce specific contexts.}
\begin{tabular}{|c|c|c|c|c|}
\hline
\textbf{Word} & \textbf{d=1}  & \textbf{d=2} & \textbf{d=3} & \textbf{d=4}  \\
\hline
sleep		&	15.74 & 0.66 & 0.00	&	0.18 \\
silence	&	20.95 & 2.82 & 0.30	&	0.36 \\
happy		&	19.92 & 1.75 & 0.05	&  0.00 \\
hair		&	26.14 & 1.31 & 0.15  &	0.00 \\
God		&	32.79 & 1.18 & 0.06  &  1.21 \\
die		&	18.20 & 2.28 & 0.30  &  0.86 \\
cold		&	9.19  & 0.69 & 0.00	&  0.60 \\
chair		&	18.31 & 1.93 & 0.40	&	0.00 \\
\hline
\end{tabular}
\end{table}

\subsection{Using the consistency index to recognize authorship}

\begin{figure}
\begin{center}
	\includegraphics[width=0.85 \textwidth]{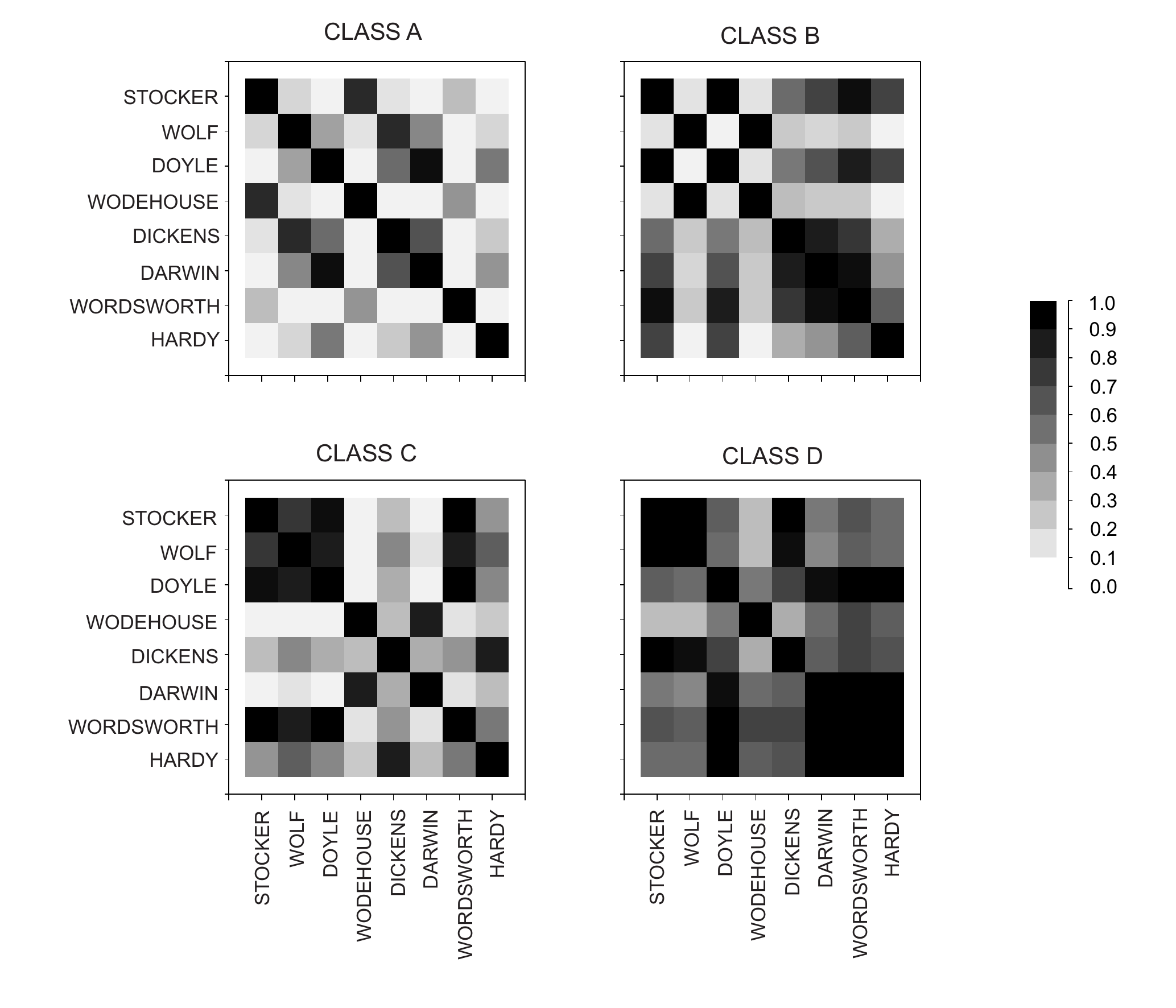}
\end{center}
\caption{Comparison of the average consistency indices for classes $C_A$, $C_B$, $C_C$ and $C_D$. The brighter the grayscale the smaller the p-value and the higher the difference in the average consistency.}
\label{fig.6}
\end{figure}

The as-expected correlations with linguistic features confirm the suitability of the consistency indices to quantify {semantic aspects} in text. We now check whether these indices can be used to identify authorship as authors  {may use words in more or less consistent ways}. In the extreme case in which one compares authors of {distinct genres (such as storytellers and authors writing scientific books)}, a very good distinction should be expected. This hypothesis was investigated by defining for each author an average consistency as:

\begin{equation} \label{cmedia}
	\overline{\mathcal{C}} = \frac{1}{n_f} \sum_{k=1}^{n_p} \mathcal{C}_k f_k,
\end{equation}
where $n_p$ and $n_f$ represent respectively the number of distinct words and the number of tokens in the corpus of a given author, $\mathcal{C}_k$ is the consistency index for the $k$-th word and $f_k$ is the frequency of the $k$-th word. Using $C_{hist}$ to compute $\overline{\mathcal{C}}$ for each author in eq. (\ref{cmedia}), we compared the average consistency of all 28 pairs of distinct authors. Figure \ref{fig.6} illustrates with grayscales the p-values from the comparison of $\overline{\mathcal{C}}$.
While some pairs of authors are easily distinguishable (see e.g. Stocker and Darwin in $C_A$, Wolf and Hardy in $C_B$, Darwin and Doyle in $C_C$ and Wodehouse and Dickens in $C_D$), others are quite similar (Stocker and Wolf in $C_D$). In addition, counting the number of darkish grayscales one notes that the words in $C_A$ provide a better ability of distinction, while class $C_D$ provides the worst.

\begin{figure}
\begin{center}
	\includegraphics[width=0.9\textwidth]{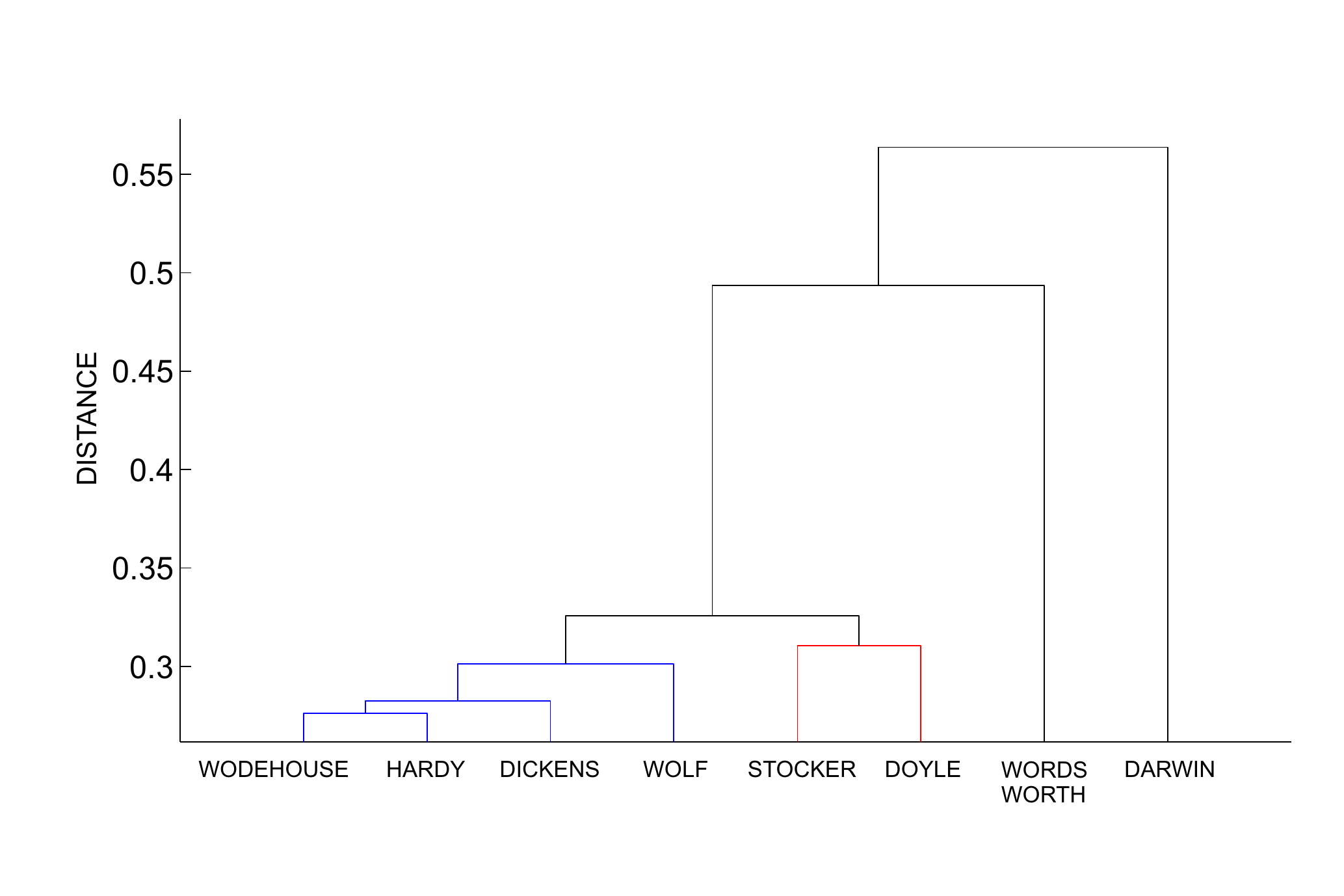}
\end{center}
\caption{Hierarchical grouping using the average consistency. Note that the texts by Darwin can be considered as outliers, since they are predominantly scientific where words should be used in a more consistent way.}
\label{fig.7}
\end{figure}

Finally, to examine how the authors are clustered in terms of consistency, we used the hierarchical clustering algorithm UPGMA~\cite{dubes} based on the cosine similarity\footnote{Each author is characterized as a vector so that each element of the vector stores the frequency of the corresponding word in the author's book. This model is widely used in text mining research and is known as bag of words~\cite{bag}.} to generate a hierarchy of authors. Figure \ref{fig.7} shows that four clusters can be identified upon choosing an appropriate distance threshold. Significantly, authors of novels (such as Wodehouse), are separated from those of scientific works (such as Darwin),{which indicates that the difference in style may be reflected in the use of words of distinct consistency indices}.

In summary, consistency indices are useful to detect authorship, which now can be combined with other conventional methods~\cite{rec1,rec0,rec2,rec3,rec4} to enhance accuracy rates in distinguishing authors.

\section{Conclusion and final remarks}

In this paper we have studied the problem of quantifying the complexity of writing considering the {consistency of words}. Assuming that {consistent words} induce grammatically/semantically limited contexts, we defined several indices to measure the tendency of a word to be used homogeneously (i.e., preserving the context). With the various indices proposed, we found that the consistency of words is well fitted by a log-normal distribution, in contrast to the Zipf's law for ranking words according to frequency. Interestingly, we found that consistency can be seen as a multiplicative growth process. Each new writer using the word modifies the current consistency by adding or removing a fraction of the current value which is independent of the current consistency. Furthermore, more frequent and familiar words tend to be used more consistently, while ambiguous words and words which take a long time to be learned tend to be less consistent. Finally, we confirmed that the quantification of complexity in terms of the characterization of the {consistency} of words is able to distinguish authors, especially those with different genres of writing.

	As future work we propose the use of new metrics for consistency in order to ascertain whether the results are preserved. As a starting point, we intend to make use of the so-called Katz similarity~\cite{katz} to quantify consistency using the distance between concepts. Similarly, we intend to further study the behavior of consistency extending the measures developed here considering not only the immediate neighborhood of the word, but also more distant neighborhoods. Since the consistency of words is related to the semantics, we wish to combine it with the features related to the writing style (for instance using CN topological measurements~\cite{costa2,rec4}) to verify if the distinguishability is enhanced. Finally, we suggest that the consistency indices may also be useful in applications to {quantify subjectivity (in written texts or in transcribed speech), for subjective words might have lower consistency values because distinct persons probably associate distinct neighbors to subjective concepts.}

\section{Acknowledgments}
This work was supported by FAPESP and CNPq (Brazil).


\end{document}